\documentclass[12pt]{article}
\usepackage{epsfig}

\def\a{\alpha}
\def\b{\beta}

\def\th{\theta}
\def\S{\Sigma}
\def\s{\sigma}
\def\f{\phi}
\def\F{\Phi}
\def\vf{\varphi}
\def\D{\Delta}
\def\e{\epsilon}
\def\r{\rho}

\def\G{\Gamma}
\def\g{\gamma}

\def\d{\delta}
\def\m{\mu}
\def\n{\nu}
\def\l{\lambda}
\def\L{\Lambda}

\def\i{\iota}
\def\h{\eta}

\def\cd{{\cal D}}
\def\cc{{\cal C}}

\def\ca{{\cal A}}

\def\ch{{\cal H}}

\def\pa{\partial}
\def\to{\rightarrow}

\newcommand{\be}{\begin{equation}}
\newcommand{\ee}{\end{equation}}
\newcommand{\bea}{\begin{eqnarray}}
\newcommand{\eea}{\end{eqnarray}}

\def\ket#1{\left|#1\right\rangle}

\begin{document}

\begin{titlepage}

\bigskip
\bigskip
\bigskip
\bigskip

\begin{center}

{\bf{\Large String Theory and Quantum Spin Networks}}

\end{center}
\bigskip
\begin{center}
 A. Mikovi\'c \footnote{E-mail address: amikovic@ulusofona.pt}
\end{center}
\begin{center}
Departamento de Matem\'atica e Ci\^{e}ncias de Computac\~ao,
Universidade Lus\'ofona, Av. do Campo Grande, 376, 1749-024,
Lisboa, Portugal
\end{center}

\normalsize

\bigskip
\bigskip
\begin{center}
                        {\bf Abstract}
\end{center}
We propose an approach to formulating string theory in a curved
spacetime, which is based on the connection between the states of
the WZW model for the isometry group of a background spacetime
metric and the representations of the corresponding quantum group.
In this approach the string states scattering amplitudes are
defined by the evaluations of the theta spin networks for the
associated quantum group. We examine the evaluations given by the
spin network invariants defined by the spin foam state sum model
associated to the two-dimensional BF theory for the background
isometry group. We show that the corresponding string amplitudes
are well-defined if the spacetime manifold is compact and admits a
group metric. We compute the simplest scattering amplitudes in the
case of the SU(2) background isometry group, and we provide
arguments that these are the amplitudes of a topological string
theory.

\bigskip
PACS: 11.25.Pm, 11.25.Hf, 0460.Pp

MSC: 81T30, 57R56, 17B37, 17B67

Keywords: $SU(2)$ WZW model, BF theory, spin foams

\end{titlepage}
\newpage

\section{Introduction}

In the string theory approach to quantum gravity one is able to
calculate perturbatively the graviton scattering amplitudes in a
flat background metric spacetime, at least up to two loops
\cite{dHP}, although there are strong indications that one can do
this to an arbitrary number of loops \cite{Ma}. From these
amplitudes one can obtain the quantum effective action for general
relativity \cite{Tsey,FJ}, and therefore in that sense the string
theory can be considered as a quantum theory of gravity. However,
beside proving the finiteness, another problem is that one is
restricted only to the spacetimes with non-compact spatial
sections, and hence addressing the problem of the
phenomenologically relevant de-Sitter spacetime becomes a
complicated task \cite{dS}.

In the canonical loop quantum gravity (CLQG) approach, for a
review and references see \cite{Smo}, the topology of the spatial
slice is not a problem, although the spacetime topology is
restricted to ${\bf R}\times\S$, where $\S$ is a three-manifold.
In this approach one is dealing with the wavefunctions of the
spatial spin connection, and therefore it is a nontrivial task to
obtain an appropriate vacuum state, which has to be peaked around
particular spacetime metric \cite{Mlgv}. A related problem is to
obtain an effective action which would correspond to general
relativity, which is related to the problem of constructing the
graviton scattering amplitudes.

The spin foam (SF) state sum models of quantum gravity, for a
review and references see \cite{sfr,nsfr}, can be considered as
the category theory generalizations of the Regge calculus approach
to quantum gravity. As such, the SF models do not have the
spacetime topology restriction as the CLQG models, although one
can think of the SF models as the path-integral generalization of
the CLQG models. All these features allow one to construct the SF
scattering amplitudes in a straightforward manner
\cite{Mqsn,Msfm,Mym}. However, because the spin foam models are
the connection based formulations, it is difficult to extract the
corresponding effective actions, and hence it is difficult to see
whether they have a good semi-classical limit, which should be the
general relativity with small quantum corrections.

Note that the string theory can be formulated as a two-dimensional
(2d) state sum model, which is known as the random triangulations
model (for a review and references see \cite{rsr}). In this
approach the string world-sheet is triangulated and embedded in a
flat metric $d$-dimensional spacetime. One then assigns a unit
length to all the edges and calculates the weights $e^{-\b S(T)}$
for each triangulation $T$, where $S(T)$ is the string action and
$\b$ is a parameter. The partition function is then obtained as a
sum of the weights over all the triangulations. An analogous
construction can be also made for the superstring \cite{gst,fst}.

By using the dual triangulation of the world-sheet, one can
represent the terms in the sum over the triangulations as the
$d$-dimensional momentum space Feynman diagrams for a $\phi^3$
matrix field theory with the $\phi e^{-\nabla^2}\phi$ kinetic
term. Unfortunately, this approach is feasible only for strings
propagating in $d \le 1$ spacetimes. The $d=0$ case can be
interpreted as a partition function for the 2d general relativity,
while the $d=1$ case can be interpreted as a partition function
for a 2d dilaton gravity with a single scalar field. Note that the
Jackiw-Taitelboim 2d dilaton gravity model can be represented as
the $SO(2,1)$ BF theory \cite{Mqsn}, so that the corresponding
spin foam model in the Euclidian case can be also written as a
$\phi^3$ matrix model \cite{Mqsn}. However, while in the random
triangulations approach one has a single matrix of a fixed
dimension, in the spin foam case one has a multi-matrix model,
with the matrices of all dimensions.

Therefore one can think of the string theory matrix models as
special cases of the BF theory 2d spin foam models. These spin
foam models can be embedded in $d>1$ spacetimes via the choice of
the BF theory Lie group $G$, so that the group manifold is the
spacetime. The spin foam state sums can be defined rigorously via
the representations of the quantum group for $G$. Since there is a
close connection between the states of the Wess-Zumino-Witten
(WZW) model for the Lie group $G$ and the representations of the
corresponding quantum group \cite{Ki,BK,FKcf}, we propose in this
paper to interpret certain spin network observables for the 2d BF
spin foam model as the scattering amplitudes for a string
propagating on the spacetime whose background metric is given by
the Lie group manifold metric. In section two we briefly describe
the construction of the string scattering amplitudes via the
vertex operators and the relation to the spacetime fields
effective action. In section three we apply the approach of the
previous section to the case of the $SU(2)$ WZW model, and by
using the connection of the WZW model to the quantum group theory,
we propose a formula for the string states scattering amplitude as
a group covariant linear combination of the $\th_n$ spin network
invariants. In section four we define these invariants via the
spin foam state-sum model based on the 2d BF theory for the Lie
group $G$, while in section five we show how this state sum can be
evaluated via the quantum group spin networks. Because our
amplitudes are based on a topological 2d theory, one can expect
that these are the amplitudes of a topological string theory, so
that we present arguments for this in section five. We present our
conclusions in section six, and in the appendix we prove the
topological invariance of our state-sum model.

\section{String theory approach}

The string theory approach to quantizing matter and gravity can be
described in the following way. Let $M$ be a $d$-dimensional
manifold representing the spacetime ($d\ge 2$), and let $\S$ be a
surface (2d compact manifold) embedded in $M$, representing the
string world-sheet. Given an embedding $X:\S \to M$, we can
associate to it a string action \bea S(X)&=&{1\over l_s^2}\int_\S
d^2 \s [ \sqrt{|g|} g^{\a\b}\pa_\a X^\m \pa_\b X^\n G_{\m\n}(X) +
\e^{\a\b}\pa_\a X^\m \pa_\b X^\n B_{\m\n}(X)\nonumber\\ &+& \vf
(X) + l_s^2 \sqrt{|g|} R\F (X) ] \quad, \eea where $l_s$ is the
string length constant, $G_{\m\n}$ is a metric on $M$, $B_{\m\n}$
a two-form on $M$, $\vf$ and $\F$ are scalar fields on $M$. The
metric $g_{\a\b}$ on $\S$ is independent from the induced metric
$\pa_\a X^\m \pa_\b X^\n G_{\m\n} (X)$, but it should belong to
the same signature class as the induced one. One can also
introduce the open string ($\S$ with boundaries) and the fermionic
string coordinates ($M$ a super-manifold), but in order to keep
the discussion simple, we will only consider the bosonic closed
string.

The quantization procedure amounts to defining the following path
integral \be Z[G,B,\vf,\F] = \int\,\cd g \,\cd X e^{iS(X)}
\quad,\ee which will be a functional of the spacetime fields
$G_{\m\n},B_{\m\n}$, $\vf$ and $\F$. In order to define $Z$ one
sets $\F = const.$ on $\S$ and splits the spacetime fields as \be
G_{\m\n} = G_{0\m\n} + h_{\m\n} \quad,\quad B_{\m\n} = B_{0\m\n} +
b_{\m\n} \quad, \ee such that the part quadratic in derivatives of
the action $S(G_0,B_0,X)$, which we denote as $S_0(X)$, is
solvable. One can then write
\bea Z[G,B,\vf] &=& \int\,\cd g \,\cd X e^{iS_0 (X)+iS(h,b,\vf,X)}\nonumber \\
&=& \int\,\cd g \,\cd X e^{iS_0 (X)}\sum_{n=0}^\infty {i^n \over n!}
S^n(h,b,\vf,X) \nonumber\\
&=& \sum_{n=0}^\infty {i^n \over n!}\int\,\cd g \,\cd X e^{iS_0 (X)}
S^n(h,b,\vf,X) \nonumber\\
&=& \sum_{n=0}^\infty {i^n \over n!}\left\langle
S^n(h,b,\vf,X)\right\rangle \quad,\eea so that the problem of
defining $Z$ is reduced to the problem of defining simpler
path-integrals.

The next step is to expand the fields $h$, $b$ and $\vf$ over a
basis of functions on $M$. This basis is chosen from the
particle-like unitary representations of the isometry group of the
metric $G_0$. For example, when $M={\bf R}^d$ and $G_0 = diag
(-,+,...,+)$ the isometry group is a d-dimensional Poincare group
$ISO(d-1,1)$, whose particle unitary irreps are labelled by a pair
$(m,s)$, where $m\ge 0$ is the mass and $s$ is an $SO(d-1)$ or
$SO(d-2)$ irrep, corresponding to the spin ($m>0$) or the helicity
of the particle ($m=0$). Since ${\bf R}^d =ISO(d-1,1)/ SO(d-1,1)$,
one takes the plane waves $u_p(X) = e^{ip_\m X^\m}$ as the basis
functions. Hence given a function $f$ on $M$ one can write \be f
(X) = \int dp\,u_p(X)\, f(p) \quad.\label{fexp}\ee In general case
$\int dp$ could be also a sum, which depends on the choice of $M$
and the background metric isometry group. In particular, when
$b=\vf=0$, the Fourier expansion (\ref{fexp}) gives \be Z[G] =
\sum_{n=0}^\infty {i^n \over n!}\int dp_1 \cdots\int dp_n
h_{\m\n}(p_1)\cdots h_{\r\s}(p_n)
A^{\m\n\cdots\r\s}(p_1,...,p_n)\,, \ee where \be
A^{\m\n\cdots\r\s}(p_1,...,p_n)=\int_\S d^2 \s_1 \cdots \int_\S
d^2 \s_n \left\langle V^{\m\n}_{p_1}(X(\s_1))\cdots V^{\r\s}_{p_n}
(X(\s_n)) \right\rangle \,,\ee and $V^{\m\n}_p (X(\s))$ is the
trace-free symmetric part of \be
 \sqrt{|g|}g^{\a\b}\pa_\a X^{\m}\pa_\b X^{\n}
u_p (X)\quad. \label{gver}\ee The tensor $V_p^{\m\n}$ represents
the graviton vertex operator, while the trace part of (\ref{gver})
gives the vertex operator for the dilaton $\f$. The antisymmetric
part of (\ref{gver}) is the B-field vertex operator. The vertex
operator $V_p = u_p (X)$ corresponds to the ground state scalar
field $\vf$, the tachyon.

The quantity $\langle V_1 \cdots V_n \rangle$ can be identified as
a correlation function of primary fields and their descendants in
a conformal field theory defined by the action $S_0$ on the
surface $\S$. The amplitude \be A(p_1,...,p_n) = \int_\S d^2
\s_1\cdots \int_\S d^2 \s_n \langle V_1 \cdots V_n
\rangle\quad,\label{scata}\ee is then interpreted as the amplitude
for scattering of $n$ quanta of the massless string states via the
world-sheet $\S$ in the spacetime $M$ with the background metric
$G_{0\m\n}$ and the background two-form $B_{0\m\n}$.

Since $\F = const.$, then the quantity $\l = e^{-i\F}$ can be
interpreted as the string theory coupling constant, so that the
contributions to the scattering amplitudes from the world-sheets
of various genra can be written as \be A_{tot}(p_1,...,p_n
)=\sum_{g\ge 0} \l^{2g-2 +n} A_g (p_1 ,..., p_n )\quad,\ee where
$g$ is the genus of the surface $\S$. In analogy to the Feynman
diagrams of particle field theories, the functional \be
Z[G,B,\f,\vf] = \sum_{g,n} \l^{2g-2 +n} Z_{g,n} [G,B,\f,\vf]
\quad,\label{strea}\ee where $n=n_G + n_B + n_\f +n_\vf$, can be
interpreted as the quantum effective action for the massless
string modes \cite{Tsey,FJ}.

One of the interesting features of string theory is that the genus
zero functional $Z_0$ can be expanded in the powers of the string
length $l_s$ as \be Z_0 [G,B,\f]  = S_{0}[G,B,\f]  + l_s^{2}S_1
[G,B,\f] + l_s^{4} S_2 [G,B,\f] + \cdots \quad,\ee where $S_0$ is
the Einstein-Hilbert action coupled to the dilaton $\f$ and the
$B$ field\footnote{Since the tachyon $\vf$ is not present in the
superstring case, we omit it here.}, while the higher-order terms
contain the higher powers of the Ricci curvature $R_{\m\n}(G)$.
The higher-genus terms in (\ref{strea}) can be then interpreted as
the quantum corrections to this classical action. This is why the
string theory can be considered as a quantum theory of general
relativity. However, it is not obvious that the amplitudes $A_g$
will be finite for every genus $g$. There are strong indications
that this can be achieved in the superstring theory \cite{Ma},
although an explicit proof has not been given yet. So far, only
the $g\le 2$ amplitudes have been proven finite \cite{dHP}.

\section{The SU(2) WZW model}

Let us analyze the $SU(2)$ WZW model in the framework of previous
section. In this case $M = S^3$ and the background metric $G_0$ is
given by \be Tr\,\int_{\S}\, d^2 \s \,\h^{\a\b} g^{-1}\pa_\a g
g^{-1}\pa_\b g \quad,\ee where $g: \S \to SU(2)$, $\h_{\a\b}$ is a
conformal class metric on $\S$, while the background field $B_0$
is given by \be Tr\,\int_{N}\, g^{-1}dg\wedge g^{-1}dg \wedge
g^{-1}dg\quad,\ee where $N$ is a compact three-manifold such that
$\pa N = \S$. The functions on $S^3$ can be expanded via the
Peter-Weyl formula for the functions on $SU(2)$ \be f(g) =
\sum_{j,m,\bar m } D^{(j)}_{m,\bar m}(g) f_j^{m,\bar m}\quad, \ee
where $2j=0,1,2,...$ and $-j\le m,\bar m \le j$. Hence the analogs
of the $u_p (X)$ functions are the $D^{(j)}_{m,\bar m}(g)$
functions where $g = e^X$. The tachyon vertex operators will be
then given by the primary fields \be V^j_{m,\bar m}(\s) =
D^{(j)}_{m,\bar m}(e^{X(\s)}) \quad,\ee which correspond to the
``particle" states $\ket{j,m}\otimes\ket{j,\bar m}\in V_j\otimes
V_j^*$ of the isometry group $SU(2)$.

Let $\S = S^1 \times\bf R$, then the first-quantized Hilbert space
of states is given by \be \ch_s = \sum_j^{\oplus} \,\hat
V_j\otimes \hat V_j^* \quad,\ee where $\hat V_j$, and its dual
$\hat V^*_j$ are the highest-weight irreps of the left and the
right Kac-Moody (KM) algebras $J_+$ and $J_-$. If we fix the level
number $k$ of the KM algebras, then there is a restriction $j\le
k/2$, because only then the corresponding irreps are unitary, i.e.
there are no negative norm states.

Note that the structure of the space $\ch_s$ can be interpreted
also as a Hilbert space of the WZW model for the coset \be
SO(4)/SO(3) = SU(2)\times SU(2)/SU(2) \quad.\ee In the particle
limit, the physical irreps are $(j,j)$, which satisfy $J_+^2 =
J_-^2$, and they are the class-one irreps with respect to the
$SU(2)$ subgroup (i.e. they contain an $SU(2)$ invariant vector,
see \cite{fksi}). One can also define a chiral WZW model Hilbert
space as \be \ch_+ = \sum_j^{\oplus} \hat V_j \quad.\ee

The correlation functions \be  \langle V^{j_1}_{m_1,\bar
m_1}(\s_1)\cdots V^{j_n}_{m,\bar m_n}(\s_n)\rangle
\quad,\label{corf}\ee can be calculated on the sphere ($\S = S^2$)
via the Knizhnik-Zamolodchikov equation \cite{KZ,ZF}, or by its
generalization when $\S$ is a higher-genus surface \cite{hKZ}. A
less rigorous approach, but more familiar to physicists, is to
reduce (\ref{corf}) to gaussian functional integrations via the
BRST gauge fixing procedure \cite{Gaw}.

As far as the graviton, dilaton and the antisymmetric field
correlation functions are concerned, it is natural to conjecture
that they would be given by the irreducible parts of the vertex
operator \be V^{ab}_{j,m,\bar m} (\s) = J_+^{a}(\s)
J_-^{b}(\s)D^{(j)}_{m,\bar m}(e^{X(\s)})\quad,\ee where $J_\pm$
are the WZW chiral currents.

In order to obtain the scattering amplitude, one would have to
integrate the correlation functions over the surface $\S$. One
expects to obtain a finite (or a renormalisible) result, because
of the topological nature of three-dimensional gravity. But when
dim$\,M> 3$, one expects to obtain non-renormalisible
diveregencies, and one would have to use a super group extension
of the isometry group in order to tame these divergencies, as it
is done in the case of flat spacetime superstring theory. However,
this program is difficult to implement, and as in the superstring
case, it is not obvious that the arbitrary genus amplitudes will
be finite. The idea of this paper is to use the connection between
the states of the WZW model and the representations of the
corresponding quantum group in order to give a more direct and
simpler definition of the scattering amplitudes $A_g$.

The quantum group connection comes from the following fact. Let
$\S = S^2$, then \be  \langle V^{j_1}_{m_1,\bar m_1}(\s_1)\cdots
V^{j_n}_{m,\bar m_n}(\s_n)\rangle = \sum_\i \psi^{j_1 \,...\,j_n
\,(\i)}_{m_1 \,...\, m_n}(z_1,...,z_n) {\bar\psi}^{j_1 \,...\,j_n
\,(\i)}_{\bar m_1 \,...\, \bar m_n}(\bar z_1,...,\bar z_n
)\quad,\label{holf}\ee where the vectors $\psi^{(\i)}$ (as well as
the $\bar\psi^{(\i)}$) form a basis in a subspace $\cc_k (j_1
,..., j_n)$ of the $\textrm{Hom}(j_1 ,..., j_n)$ space, where $k$
is the level. The vector space $\cc_k$ is known as the space of
conformal blocks, and $\cc_k$ is isomorphic to the $\textrm{Hom}_q
(j_1 ,..., j_n)$ space for the irreps of the quantum group $U_q
(su(2))$ for $q=\exp\left({i\pi\over k+2}\right)$. More generally,
if $\bf{g}$ is a simple Lie algebra, then the category of the
integrable irreducible representations (irreps) of the affine Lie
algebra $\hat{\bf{g}}$, based on the finite-dimensional irreps of
$\bf{g}$, at the level $k\in \bf{Z}_+$ is equivalent as a modular
tensor category to the category of the finite-dimensional irreps
of the quantum group $U_q(\bf{g})$ for $q=\exp{i\pi\over m(k+h)}$,
where $m\in \bf{N}$ and $h$ is the dual Coexter number \cite{BK}.
In the case of Lie algebras $sl(n)$ or $su(n)$, $m=1$ and $h=n$.

One can then write \be \psi^{j_1 \,...\,j_n \,(\i)}_{m_1 \,...\,
m_n}(z) = v_{\i}^{(k)}(z_1,...,z_n )C^{j_1 \,...\,j_n \,(\i)}_{m_1
\,...\, m_n} \quad,\ee where $C^{(\i)}$ is the intertwiner tensor
from the Hom$(j_1,...,j_n)$ space, and similarly for $\bar\psi$,
so that the scattering amplitude will be given as \be A_{m_1 \bar
m_1 ... m_n \bar m_n}^{\,\,\,\,j_1 \,\,\,\,\, ...\,\,\,\,\, j_n} =
\sum_\i N_\i (j_1,...,j_n |k)C^{j_1 \,...\,j_n \,(\i)}_{m_1
\,...\, m_n} \left( C^{j_1 \,...\,j_n \,(\i)}_{\bar m_1 \,...\,
\bar m_n}\right)^* \quad,\label{spa}\ee where \be N_\i
(j_1,...,j_n |k)= \int dz_1\, d\bar z_1 \cdots \int dz_n\, d\bar
z_n \, v_\i^{(k)} (z) \bar v_\i^{(k)} (\bar z )\quad.\ee

The constant $N_\i$ can be identified as a norm of the vector
$v_\i$ from the $\cc_k$ space, and the value of this norm can be
related to an evaluation of the $\th_n$ spin network
\cite{Ki,FKcf}. On the sphere, this evaluation is determined up to
a numerical constant, i.e. one can have many different
evaluations. We will take a particular evaluation, which is
suitable for our purposes, given by \be N_\i (j_1,...,j_n |k) =
\langle \theta (j_1,...,j_n |\i)\rangle_{q=\exp{i\pi\over
k+2}}\quad, \label{qsn}\ee where $\langle\th\rangle_q$ is the
evaluation defined by a 2d spin foam state sum invariant for the
theta spin network embedded in a triangulation of $\S$. This is
motivated by the results from \cite{BD,FKcf,Mlgv} on the relation
between the spin networks, quantum groups and the state-sum
models.

Note that the two intertwiners for the $\th_n$ spin network can be
different in general, and hence one should specify \be N_{\i
\i^{\prime}} (j_1,...,j_n |k) = \langle \theta (j_1,...,j_n
|\i,\i^{\prime})\rangle_{q=\exp{i\pi\over k+2}}\quad.
\label{gqsn}\ee Therefore the string amplitude will be given by
\be A_{m_1 \bar m_1 ... m_n \bar m_n}^{\,\,\,\,j_1 \,\,\,\,\,
...\,\,\,\,\, j_n} = \sum_{\i \i^{\prime}} N_{\i\i^{\prime}}
(j_1,...,j_n |k)C^{j_1 \,...\,j_n \,(\i)}_{m_1 \,...\, m_n} \left(
C^{j_1 \,...\,j_n \,(\i^{\prime})}_{\bar m_1 \,...\, \bar
m_n}\right)^* \quad.\label{gspa}\ee

\section{String spin-foam model}

The considerations of the previous sections suggest that one
should construct a theta spin network evaluation which would
depend on the string world-sheet, i.e. the surface $\S$, and its
embedding into a spacetime, i.e. the group manifold $G$. This
evaluation should be a 2d topological invariant, because of the
invariance under the world-sheet reparametrizations
(diffeomorphisms of $\S$). A natural way to obtain this invariant
is to take the 2d BF theory for the Lie group $G$, and consider
the spin network observable \be \langle\th(j_1 ,...,j_n
;\i,\i^{\prime})\rangle = \int \cd\,B\,\cd\,A \,e^{i\,Tr\int_\S
B\wedge F} \th(j_1 ,...,j_n ;\i,\i^{\prime} |A)
\quad,\label{thpi}\ee where $\th(j_1,...,j_n;\i|A)$ is the spin
network function associated to the $\th_n$ spin network, i.e. a
product of the holonomies along the edges of the spin network
contracted by the intertwiners $\i$ and $\i^{\prime}$ \cite{sfr}.

Note that one can construct other 2d diffeomorphism invariant
path-integral expressions based on the BF theory. One can also
take a non-topological modification of the BF theory, and then sum
over the triangulations in order to make a diffeomorphism
invariant. We have taken a simple choice, based on a topological
theory, so that one would not need to perform a sum over the
triangulations. However, we will see that the amplitudes one
obtains from (\ref{thpi}) are those of a topological string
theory.

The expression (\ref{thpi}) is a generalization of the expectation
value of the Wilson loop, and in order to get an invariant, we
need to define this path-integral. This can be done by integrating
the $B$ field, which gives a $\d (F)$, and then triangulating the
$\S$, so that the $A$ integration is replaced by the integration
over the group elements (holonomies) associated to the dual
lattice links \cite{fksf,sfr,pfo,Pf}. The spin network function
$\th(A)$ is then given by a product of the corresponding
representation matrix group elements (associated to the edges of
the spin network) contracted by the appropriate intertwiner
tensors (associated to the vertices of the spin network). The
group integrations can be performed by using the following group
theory formulas \be \d (g) = \sum_{\L} \textrm{dim}\,\L
\,\chi_{\L} (g)\quad, \ee and \bea \int_G \,dg
\,D^{(\L_1)}_{\a_1\b_1} (g) D^{(\L_2)}_{\a_2\b_2} (g) &=&
{1\over\dim\L_1}\d_{\L_1 ,\L_2} C^{{\L_1}{\L_2}}_{\a_1 \a_2}
\left(C^{{\L_1}{\L_2}}_{\b_1 \b_2}\right)^*
\\\int_G \,dg \,D^{(\L_1)}_{\a_1\b_1} (g) \cdots D^{(\L_n)}_{\a_n\b_n} (g) &=&
\sum_{\iota} C^{{\L_1}\cdots {\L_n}(\iota)}_{\a_1 \cdots \a_n}
\left(C^{{\L_1}\cdots {\L_n}(\iota)}_{\b_1 \cdots \b_n}\right)^*
\,,\, n\ge 3\,,\label{edgi}\eea where $D^{(\L)}$ is the group
representation matrix in the representation $\L$, and $\chi_\L$ is
the corresponding trace. One then obtains \be \langle\th_n
(j,\i,\i^{\prime})\rangle = \sum_{\L_f ,\i_e}\ca
(\L_f,\i_e,j_1,...,j_n,\i,\i^{\prime} )\label{sfs}\ee where $\L_f$
are the irreps associated to the faces of the dual two-complex for
$\S$, $\i_e$ are the edge intertwiners coming from (\ref{edgi})
and $\ca$ is the amplitude for such a colored two-complex. This
sum is called the spin foam state sum, because the faces of the
two-complex remind us of a soap foam, and coloring by the irreps,
i.e. the spins, gives a spin foam.

The amplitude $\ca$ turns out to be a product of the group theory
spin network evaluations, and this form of the amplitude can be
obtained in the following way. Let $\G$ be the one-complex for a
dual triangulation of $\S$, i.e. $\G$ is a trivalent graph. We can
thicken $\G$ by replacing the edges by the ribbons, and we denote
the corresponding ribbon graph as $\tilde\G$, see Fig. 1. Let us
color every face (closed loop of $\tilde\G$) by the irreps
$l_1,...,l_L$. On the thickened graph draw the theta graph, such
that the vertices of the theta graph coincide with the vertices of
$\G$, and the edges of the theta graph run along the edges of
$\tilde\G$, see Fig. 2. Then label the edges of the theta graph by
the irreps $j_1,...,j_n$, as well as the two vertices of theta
graph by the intertwiners $\i$ and $\i^{\prime}$. In this way we
obtain a colored $\tilde\G_c$ graph, see Fig. 2. Now shrink the
middle of each edge of $\tilde\G_c$ to a point. In this way we
obtain a collection of spin networks, whose number is equal to the
number of the vertices of $\G$, see Fig. 3. To this configuration
of spin networks we associate the following amplitude \be \ca
=\prod_f \textrm{dim}(l_f)\prod_e A_e (l_f,j)\prod_v A_v
(l_f,\i_{e},j,\i,\i^{\prime}) \quad. \label{csfa}\ee The edge
amplitude $A_e$ is given by ${\dim}^{-1}(l_{f(e)})$ if none of the
theta graph edges pass through the ribbon edge $e$, otherwise it
is $1$. The vertex amplitude $A_v$ is given by the group theory
spin network evaluation for the spin network at the vertex $v$.

The group theory evaluation of a spin network is given by the
product of the vertex intertwiner tensors $C^{(\i)}$, which are
contracted by the $\d_\a^\b$, $C_{\a\b}$ or $(C_{\a\b})^* =
C^{\a\b}$ tensors associated to the edges. For example, the
evaluation of the $\th_3$ spin network is given by
\be\th_3(\L_1,\L_2,\L_3) = \sum_{\a,\b,\g}
C_{\,\a\,\,\b\,\,\g}^{\L_1 \L_2
\L_3}\left(C_{\,\a\,\,\b\,\,\g}^{\L_1 \L_2 \L_3} \right)^* \quad,
\label{tht}\ee where $C$ are the normalized Clebsch-Gordan (CG)
coefficients and \be (C_{\a\b\g})^* =
C^{\a\m}C^{\b\n}C^{\g\r}C_{\m\n\r}=C^{\a\b\g} \quad.\label{cgc}\ee
In the formula (\ref{cgc}) we have suppressed the $\L$ indices on
the CG tensors for simplicity. The normalization is such that when
$\th_3$ is non-zero, its value is one.

In the case of the $\th_4$ spin network we have \be
\th_4(\L_1,\L_2,\L_3,\L_4;\L,\L^{\prime}) = \sum_{\a,\b,\g,\d}
C_{\,\a\,\,\,\b\,\,\,\g\,\,\,\d}^{\L_1 \L_2 \L_3 \L_4 (\L)}
\left(C_{\,\a\,\,\,\b\,\,\,\g\,\,\,\d}^{\L_1 \L_2 \L_3 \L_4
(\L^{\prime})}\right)^*\quad, \label{thf}\ee where \be
C_{\,\a\,\,\,\b\,\,\,\g\,\,\,\d}^{\L_1 \L_2 \L_3 \L_4 (\L)} =
\sum_{\r\m}C_{\,\a\,\,\b\,\,\r}^{\L_1 \L_2 \L }C^{\r\m}
C_{\,\m\,\,\g\,\,\d}^{\L \L_3 \L_4} \quad,\ee while the evaluation
for the tetrahedral spin network is given by a normalized $6j$
symbol \be Tetr(\L_1,...,\L_6)=\sum_\a C^{\L_1 \L_ 2 \L_ 3}_{\a_1
\a_ 2 \a_3 }C^{\L_3 \L_4 \L_6}_{\a_3^{\prime} \a_4 \a_6}C^{\L_1
\L_ 5 \L_ 6}_{\a_1^{\prime} \a_5 \a_6^{\prime}}C^{\L_2 \L_4
\L_5}_{\a_2^{\prime} \a_4^{\prime}
\a_5^{\prime}}C^{\a_1\a_1^{\prime}}\cdots
C^{\a_6\a_6^{\prime}}\quad. \label{tetr}\ee

Let us now calculate the spin foam amplitude for a Wilson loop (a
$\th_2$ spin network) on the sphere. A triangulation of the sphere
by four triangles can be represented by the Mercedes-Bentz (MB)
graph, whose ribbon version is shown in Fig. 1. The corresponding
SF amplitude will be a product of
 \be \prod_f A_f =\dim^3 l_1 \dim l_2 \,,\, \prod_e A_e =
\dim^{-3} l_1 \,,\, \prod_v A_v =\dim l_1 (\th_3 (l_1,j,l_2))^3
\,,\ee which can be seen from the Figs 2 and 3 by setting $j_1 =
j_2=0$, $j_3 =j_4 =j$ and $l_1=l_3=l_4$. Therefore \be
\langle\th_2 (j)\rangle =\sum_{l_f}\prod_f A_f \prod_e A_e \prod_v
A_v =\sum_{l_1,l_2}\dim l_1 \dim l_2 (\th_3 (l_1,j,l_2))^3
\quad.\ee Since $\th_3^2 (a,b,c)= \th_3 (a,b,c)$ and $d_a d_b =
\sum_c N_{ab}^c d_c$ where $\th_3 (a,b,c) = N_{ab}^c$ and $d_a =
\dim\,a$, we get \be \langle\th_2 (j)\rangle = d_j \sum_{l} d_l^2
\quad.\label{cwl}\ee

The sum (\ref{cwl}) is divergent and that will be a generic
problem with the spin foam state sum (\ref{sfs}). A topologically
invariant way to regularize it is to use the fact that the spin
network evaluations can be interpreted as traces of $\bf C$-linear
morphisms in the tensor category $Cat(G)$ of the irreps of the Lie
group $G$. Then the spin foam state sum $\langle\th_n\rangle$ can
be understood as a trace of a functor defined by the spin network
morphisms associated to the spin foam, which maps the $Cat(G)$ to
itself. This functor is topologically invariant because the spin
network morphisms satisfy relations associated with the invariance
under the Pachner moves (see the Appendix).

The irreps of the quantum group $U_q (\bf{g})$ also form a tensor
category $Cat_q (G)$, and hence one can define the q-spin net
evaluations as the traces of $\bf C$-linear morphisms defined by
the same spin network graphs as in the $Cat(G)$ case. As a result,
the quantum spin network evaluations are given essentially by the
same expressions as in the classical case, i.e. one replaces the
intertwiner tensors $C^{(\i)}$ by their quantum analogs, but one
must take into account the over and the under-crossings, which
give a non-trivial contribution in the quantum group case
\cite{BD,Ca,NR}. Therefore the quantum spin networks inherit the
topological properties of the classical ones. When $q$ is an
appropriate root of unity, then one obtains a modular tensor
category, which has a finite number of the irreducible objects.
One can then define the analogous functor as in the classical
case, which is now finite, and topologically invariant.

Hence let us define the invariant $N_{\i\i^{\prime}}$ as the state
sum \be \langle\th_n (j,\i,\i^{\prime})\rangle = \sum_{l_f
,\i_e}\prod_f \D (l_f)\prod_e A_e^{(q)} (l_f,j)\prod_v A_v^{(q)}
(l_f,\i_e,j,\i,\i^{\prime}) \quad,\label{qsna}\ee where $\D (l)$
is the quantum dimension and $A^{(q)}$ are the quantum evaluations
of the spin networks appearing in the $Cat(G)$ amplitude
(\ref{csfa}). In the $SU(2)$ case, one can calculate the quantum
spin networks evaluations via the representation theory of the
$U_q(sl(2))$ for $q$ a root of unity \cite{Ca}, and by using the
fact that any spin network can be represented as the trace of a
composition of the 3-morphisms (i.e. CG tensors) in the
corresponding tensor category.

As far as the gravitons and the other excited states are
concerned, to the best of our knowledge, we do not know of any
result in the literature which explores the connection between the
corresponding affine Lie algebra representation states and the
related quantum group representation states. A reasonable guess,
based on the nature of our construction, is that the scattering of
the excited states carrying the spins $s_i$ and the vacuum
representations $j_i$, where $i=1,2,...,n$, would be given by the
expectation value of the $\th_n (j)$ graph with the $s_i$ loops
linked with the $j_i$ lines, see Fig. 9. This expectation value
can be defined via the spin foam state sum for the corresponding
graph drawn on the ribbon graph for the surface, see Fig. 9. The
corresponding spin foam amplitude will have the same form as the
amplitude in the state sum (\ref{qsna}), but now the edge
amplitude $A_e$ will be different when one or more $s$-loops sit
on that edge. In that case the amplitude $A_e$ will be given by
the corresponding evaluation of the theta spin network containing
the $s_i$ links.

\section{Calculation of the invariants}

It is not difficult to show that in the case when $\S=S^2$, the
invariant $I$ is proportional to the usual q-spin network
evaluation of the theta graph. The expressions for the $A$
amplitudes are the same as in the classical case, and the only
difference is that the spin network evaluations are the quantum
ones. In the $n=2$ case (i.e. the Wilson loop), this can be seen
from the formula (\ref{cwl}), since in the quantum group case the
sum $\sum_l d_l^2$ is finite. In the $SU_q(2)$ case for
$q=e^{i\pi/k+2}$ one has $2l=0,1,2,...,k$ and the quantum Wilson
loop evaluation (or the quantum dimension) is given by \be \D_l =
(-1)^{2l}\, {\sin{\pi(2l+1)\over k+2} \over \sin{\pi\over
k+2}}\quad,\ee so that \be \sum_l \D_l^2 = {k+2\over
2\left(\sin{\pi\over k+2}\right)^2}= c_k^2 \quad.\ee Therefore \be
\langle\th_2 (j)\rangle_{g=0} = c_k^2 \D_j \quad.\ee

In the $n=3$ case, we use the topological invariance of the sum,
and instead of using the MB graph, which corresponds to
triangulating a sphere with four triangles, we can triangulate the
sphere with only two triangles. The dual graph is the theta graph,
see Fig. 4, and we obtain \bea \prod_f A_f &=& \D_a \D_b \D_c
\quad,\quad \prod_e A_e = 1 \quad,\\\quad \prod_v A_v &=&
Tetr(j_1,j_3,j_2,b,a,c)\,Tetr(j_1,j_2,j_3,b,c,a)
 \quad,\eea
so that \be \langle\th_3\rangle_0 = \sum_{a,b,c}\D_a \D_b \D_c
Tetr(j_1,j_3,j_2,b,a,c)\,Tetr(j_1,j_2,j_3,b,c,a)
\quad.\label{qtht}\ee Due to ortogonality of the normalized $6j$
symbols \cite{BD}\be \sum_l \D_l \,Tetr(i,j,l,a,b,c) \, Tetr^*
(i,j,l,a,b,c^\prime) = \D_c^{-1}\d_{c,c^\prime} N_{ic}^b N_{ja}^c
\quad,\ee and the symmetry under the permutations of the columns
of the $6j$ symbol, the sum (\ref{qtht}) becomes \be
\langle\th_3\rangle_{g=0} = c_k^2 \,\th_3 (j_1,j_2,j_3) = c_k^2
N_{j_1 j_2}^{j_3}\quad. \label{tts}\ee

By induction one can show that \be \langle\th_n
(j,\i,\i^{\prime})\rangle_{g=0} = c_k^2 \, \th_n
(j,\i,\i^{\prime}) \quad,\label{ethn}\ee where $\th_n$ is a
rescaled quantum evaluation of the $\th_n$ spin network, given by
(A.3). Clearly the $\th_n$ spin network expectation value is
independent of the way we have embed the $\th_n$ spin network in
the ribbon graph, which is the consequence of the topological
invariance of the state sum $\langle\th_n\rangle$ (see the
Appendix).

The result (\ref{ethn}) implies that the corresponding scattering
amplitudes, given by the formula (\ref{gspa}), are the amplitudes
of a topological string theory. One can see this by taking
$G=U(1)^N$ and $q=1$, so that the irrep labels become the discrete
momenta, while the Clebsch-Gordon coefficients become the delta
functions for the momenta meeting at a three-vertex. One can then
calculate the four-point function $A(p_1,...,p_4)$ for the sphere,
and one obtains \be A(p_1,...,p_4)= \,const.\, \d(p_1 +p_2 -p_3
-p_4) \quad,\label{tta}\ee where the constant is infinite. It is
plausible to assume that (\ref{tta}) will be the dominant term for
$q$ close to one, with the constant being a large finite number,
which gives a topological theory amplitude. In the non-topological
case, the constant in (\ref{tta}) will be replaced by a function
of the momenta. For example, in the usual string theory, this
function is given by the Koba-Nielsen amplitude $\G
(1+s/2)\G(1+t/2)$, where $s=(p_1+p_2)^2$ and $t=(p_2+p_3)^2$.

In the case when $\S$ is a torus, we can triangularize it by two
triangles. The dual graph is then the theta graph with two edges
over-crossed. The corresponding ribbon graph is the theta graph
with each edge twisted, see Fig. 5. When we draw a Wilson loop on
that ribbon graph, the corresponding spin-foam amplitude will be
the product of \be \prod_f A_f = \D_l  \quad,\quad \prod_e A_e =
\D_l^{-1} \quad,\quad \prod_v A_v =  \th_3 (l,j,l) \th_3 (l,j,l)
 \quad,\ee
which gives \be \langle\th_2 (j)\rangle_{g=1} = \sum_l N_{j\,
l}^{\,\,l}\quad. \label{wlt}\ee In the $\th_3$ case we get \be
\langle\th_3 (j)\rangle_{g=1} = \sum_l \D_l \left(
Tetr(j_1,j_2,j_3,l,l,l)\right)^2\quad, \label{thto}\ee while in
the $\th_4$ case, we get \bea \langle\th
(j_1,j_2,j_3,j_4;i,j)\rangle_{g=1}&=& \sum_{l,p} \D_l \, Prism
(j_1,j_4,i,j_2,j_3;l,l,p,l)\nonumber\\&&\quad\quad
Prism(j_4,j_1,j,j_3,j_2;l,l,p,l)\,,\eea where $Prism$ is the
evaluation of the spin network shown in Fig. 6.

In the case of a higher-genus surface we can use a simple
triangulation associated to the decomposition of the genus $g$
surface into $g$ tori connected by $g-1$ tubes. A triangulation of
a torus with a hole can be represented by a twisted theta ribbon
graph with a hole cut around one of the vertices and three ribbons
attached to the border of that cut, see Fig. 7. By joining these
ribbon graphs along the free ribbon edges, one obtains a trivalent
ribbon graph whose Euler characteristic $V-E+L$ is $2-2g$, where
$V$ is number of vertices, $E$ is the number of edges and $L$ is
the number of loops. For example, in the $g=2$ case one obtains a
trivalent ribbon graph with $14$ vertices, $21$ edges and $5$
loops, see Fig. 8.

As far as the amplitudes for the excited states are concerned, let
us analyze the case of a $\th_3$ spin network linked with three
loops carrying the irreps $s_1$, $s_2$ and $s_3$. On the sphere,
see Fig. 9, we will have \bea \langle\tilde\th_3(j,s)\rangle &=&
\sum_{a,b,c}
\D_a \D_b \D_c \, Tetr(1,3,2,b,a,c)\, Tetr(1,2,3,b,c,a)\nonumber\\
&&\quad\tilde\th_3(c,(j_1,s_1),a)\tilde\th_3(a,(j_2,s_2),b)\tilde\th_3(b,(j_3,s_3),c)
\quad.\eea Since \be \tilde\th_3(i,(j,s),l) = \th_3 (i,j,l)\,
{Hopf(j,s)\over \D_j} \quad,\ee where $Hopf$ is the evaluation of
two linked Wilson loops, we obtain \be
\langle\tilde\th_3(j,s)\rangle = c_k^2 \,N_{j_1,j_2}^{j_3}\,
\prod_{i=1}^3 {Hopf(j_i,s_i)\over \D_i}\quad.\ee In terms of the
modular S-matrix elements, which can be defined as \be
S_{j\,l}={Hopf(j,l)\over c_k} \quad,\ee we obtain \be
\langle\tilde\th_3(j,s)\rangle = c_k^5 \, N_{j_1,j_2}^{j_3}\,
\prod_{i=1}^3 {S_{j_i\,s_i}\over \D_i}\quad.\ee In the $SU(2)$
case these expressions can be calculated by using the formula\be
S_{j\,l} = \sqrt{2\over k+2} \sin {(2j+1)(2l+1)\pi\over k+2}
\quad.\ee

\section{Conclusions}

We have proposed to define the $n$-point scattering amplitude for
a string theory in a curved spacetime which is determined by the
group manifold of a Lie group $G$, as a linear combination of
expectation values of the $\th_n$ spin network (\ref{gspa}). We
have defined the $\th_n$ expectation values by the state sum
(\ref{qsna}) for the 2d spin foam model based on the quantum group
$G_q$, where $q$ is an appropriate root of unity. This definition
is very natural when one considers the scattering of string ground
states, due to the known connection between the space of conformal
blocks for the WZW model for the group $G$ at the level $k$ and
the corresponding Hom spaces for the $G_q$ irreps for $q=\exp{\pi
i\over k + h}$, where $h$ is the dual Coexter number. However, one
can argue that this definition gives a topological string theory
scattering amplitudes, see the eq. (\ref{tta}). Therefore a
further work is necessary to modify the path-integral expression
(\ref{thpi}) in order to obtain a non-topological string theory
amplitudes.

The ground states scattering amplitudes based on the expectation
value of the $\th_n$ spin networks can be naturally extended to
the case of scattering of the excited states with spins (irreps)
$s_i$, $1\le i\le n$, by considering the $\th_n$ spin network
linked with the loops $s_i$. However, in this case the precise
connection between the states of the $\hat{\bf{g}}$ integrable
irrep and the corresponding quantum group $U_q (\bf{g})$ irrep is
not understood, so that this issue would require more study in
order to find a correct formulation for the excited states.

Note that $G$ can be also a non-compact group, as long as we use
the category of finite-dimensional irreps. However, the
particle-like representations used in string theory are unitary,
and hence infinite-dimensional. In this case the complication is
that the spin net evaluations like (\ref{tht}), (\ref{thf}) and
(\ref{tetr}) become infinite sums or improper integrals, and
therefore there is no guarantee that they will be convergent.
However, the work on the Lorentzian spin foam models has
demonstrated that a large class of such spin network evaluations
can be defined, i.e. the class of simple spin networks
\cite{bcl,bb}. Furthermore, one can pass to the quantum group
representations, in which case the convergence properties are
enhanced \cite{NR}. Note that in the $SO(d-1,2)$ case, one can
construct unitary finite-dimensional irreps at roots of unity
\cite{stein}, so that all the spin network evaluations will be
defined in that case. In the Lorentzian case, the state sum
integral over the continuous parameter of the unitary irreps
becomes an integral over a compact interval when $q$ is real,
while it is expected that the root of unity irreps will have the
discrete parameter truncated \cite{bcl}, just as in the compact
group case. Hence there are strong indications that the spin-foam
string amplitudes could be constructed in the case of unitary
irreps for non-compact groups.

As far as the flat backgrounds are concerned, the relevant groups
are $G=U(1)^N$ or $G={\bf R}^N$ in the non-compact case, and one
can explore these examples before going to the more difficult
non-abelian cases, especially when the properties of the
corresponding quantum torii and the quantum planes have been
analyzed in the literature, see for example \cite{abqg}.

We have formulated our string model only for the spacetimes which
are group manifolds. Although this set of manifolds contains a lot
of physically relevant ones, for example $$SO(2,1)=AdS_3=S^1
\times {\bf{R}}^2 \,\,,\,\, SL(2,{\bf{R}})={\bf{R}}^3 \,\,,\,\,
SL(2,{\bf{C}})={\bf{R}}^3 \times S^3 \,\,,$$ etc., one would like
to find an analogous construction for the case of coset manifolds
$G/H$ since the Minkowski, de Sitter and anti-de Sitter spaces,
given respectively by
$$Mink_d = {ISO(d-1,1)\over SO(d-1,1)}\,\,,\,\, AdS_d =
{SO(d-1,2)\over SO(d-1,1)}\,\,,\,\, dS_d = {SO(d,1)\over
SO(d-1,1)}\,,$$ are even more important examples of such spaces.
The work on the higher-dimensional spin foam models revealed that
in the case of $G/H$ homogeneous spaces one can use the simple
spin network evaluations \cite{bce,fksi,bcl}, but the problem is
that the corresponding state sums are not topologically invariant.
In the 2d case the topological invariance means invariance under
the world-sheet reparametrizations, and the only way to restore it
would be to sum over the different triangulations of the surface
$\S$, which is in general a difficult thing to do. Perhaps using
the quantum simple spin networks \cite{Ye,NR} could improve the
convergence of that sum.

As far as the supersymmetry is concerned, it can be implemented
straightforwardly by replacing the spacetime isometry group by an
supergroup extension, for example $SO(N)\to OSp(N,2M)$.

Inclusion of boundaries in our formalism will be an important
task, since this will be a way to treat the open strings and the
D-branes. This could provide an interpretation in terms of quantum
spin networks for the amplitudes of the topological strings
propagating on Calabi-Yau manifolds \cite{akmv}.

\bigskip
\bigskip
\noindent{\bf ACKNOWLEDGEMENTS}

\bigskip
I would like to thank J. Prata, R. Picken, L. Smolin, N. Berkovits
and W. Siegel for the discussions. I would like to thank the
organizers of the 2d gravity workshop at the Erwin Schroedinger
Institute (Vienna, 15 Sep. - 12 Oct. 2003) for inviting me to talk
about this work, and for the discussions with H. Steinacker and I.
Kostov I had there. Work supported by the FCT/FEDER grants
POCTI/FNU/49543/2002 and POCTI/MAT/45306/2002.

\newpage
\noindent{\bf APPENDIX A}

\bigskip
Let us demonstrate the topological invariance of the $\th_n$ spin
network state sum (\ref{qsna}). First, we consider the case when
there is no spin network. Then the state sum (\ref{qsna}) becomes
the partition function for the 2d BF theory. It is given by $$ Z_g
= \sum_{\L} \D_{\L}^{L-E+V} =\sum_{\L}
\D_{\L}^{2-2g}\quad,\eqno(A.1)$$ which is obviously a topological
invariant. However, it will be instructive to prove this by
proving the invariance of the state sum under the 2d Pachner moves
\cite{Pa} by examining how the corresponding amplitude changes.

The invariance under the $(2,2)$ move can be represented
graphically by Fig. 10, which is equivalent to $$ \D_a {1\over
\D_a} \D_a = \D_a {1\over \D_a} \D_a \quad.$$ The invariance under
the $(1,3)$ move can be represented graphically by Fig. 11, which
is equivalent to
$$ \D_a = \D_a \D_a^{-3} \D_a^3 \quad.$$ Note that the same
equations are obtained when a $\th_n$ spin network is embedded in
$\S$, if the Pachner moves act on the part of the surface graph
where no $\th_n$ edge passes.

Now let us analyze the action of the Pachner moves on a region of
the surface graph where only one edge of the $\th_n$ graph passes.
The invariance under the $(2,2)$ move can be represented by Fig.
12, which is equivalent to $$ \th_3 (a,j,b) \th_3 (a,j,b) =
\D_a^{-1} \D_a \th_3 (a,j,b) \quad.$$ This is an identity because
of $\th_3^2 = \th_3$. The invariance under the $(1,3)$ move can be
represented graphically in two inequivalent ways, see Fig.13 and
Fig. 14. The Fig. 13 implies $$ \th_3 (a,j,b) = \D_a \D_a^{-2}
\D_a \th_3 (a,j,b) \th_3 (a,j,b)\quad,$$ while the Fig. 14 gives
$$ \th_3 (a,j,b) = \D_b \D_b^{-1} \th_3 (a,j,b) \th_3 (a,j,b)\th_3
(a,j,b)\quad.$$ These are again identities due to $\th_3^2
=\th_3$.

Now let us assume that the topological invariance holds for a
region containing $m\ge 1$ edges of a $\th_n$ spin network. We
will prove that this implies the topological invariance for a
region containing $m+1$ edges. In order to prove this, let us
first prove the following lema:

{\bf L1} Given a region with $m$ edges one can remove one edge
from that region without changing the state sum.

Proof: It is sufficient to prove the invariance of the state sum
under the elementary move represented in Fig. 15, since any larger
move of a $\th_n$ edge can be represented as a composition of the
elementary moves and their mirror images. The Fig. 15 implies the
identity represented in Fig. 16. Note that the quantum evaluation
of the $\th_n$ spin network is given by \cite{Ca}
$$ \th (j_1,...,j_n ;I,L)= {\th(j_1,j_2,i_1)\th(i_1,j_3,i_2)\cdots
\th(i_{n-3},j_{n-1},j_n)\over
\D_{i_1}\D_{i_2}\cdots\D_{i_{n-3}}}\,\d_{I,L} \,,\eqno(A.2)$$
where $I=(i_1,...,i_{n-3})$, $L=(l_1,...,l_{n-3})$ and $n\ge 4$.
The identity in Fig. 16 is satisfied if
$$\th(j_1,...,j_n ;I,L)=\th(j_1,j_2,i_1)\th(i_1,j_3,i_2)\cdots
\th(i_{n-3},j_{n-1},j_n)\,\d_{I,L}\quad.\eqno(A.3)$$ Since the
evaluation (A.3) is a rescaled quantum evaluation (A.2), this
means that the vertex amplitude for the $\th_n$ spin network must
be rescaled as
$$ \th(j_1,...,j_n ;I,L) \to \D_{i_1}\cdots\D_{i_{n-3}}
\,\th(j_1,...,j_n ;I,L) \quad,\eqno(A.4)$$ in order to have an
invariant state sum.

Now, given a region with $m+1$ $\th_n$ edges, by the lema {\bf L1}
one can remove one $\th_n$ edge from that region without changing
the state sum. By the assumption, any region containing $m$
$\th_n$ edges is invariant under the Pachner moves, and hence the
larger region which contains the $m$ edges plus the one which was
moved will be invariant under the Pachner moves.

In order to complete the proof, one should also demonstrate the
invariance under the movement of the vertices of the $\th_n$ spin
network. This is easy to establish by examining the state-sum
invariance under the movement of one vertex of the $\th_n$ spin
network from a vertex $v_1$ of the ribbon graph to a vertex $v_2$
connected by a ribbon edge. This is equivalent to an equation that
the evaluation of the vertex spin network at $v_1$ contracted by
the $\th_{m(v_1)}$ evaluation is the same as the evaluation of the
vertex spin network at $v_2$ contracted by the $\th_{m(v_2)}$
evaluation. Since we use the rescaled evaluation for $\th_m$,
given by the eq. (A.3), which is essentially the delta function of
the intertwiners, it is easy to verify the equation for the
$\th_n$ vertex movement. QED

\newpage

\begin{figure}[h]
\centerline{\psfig{figure=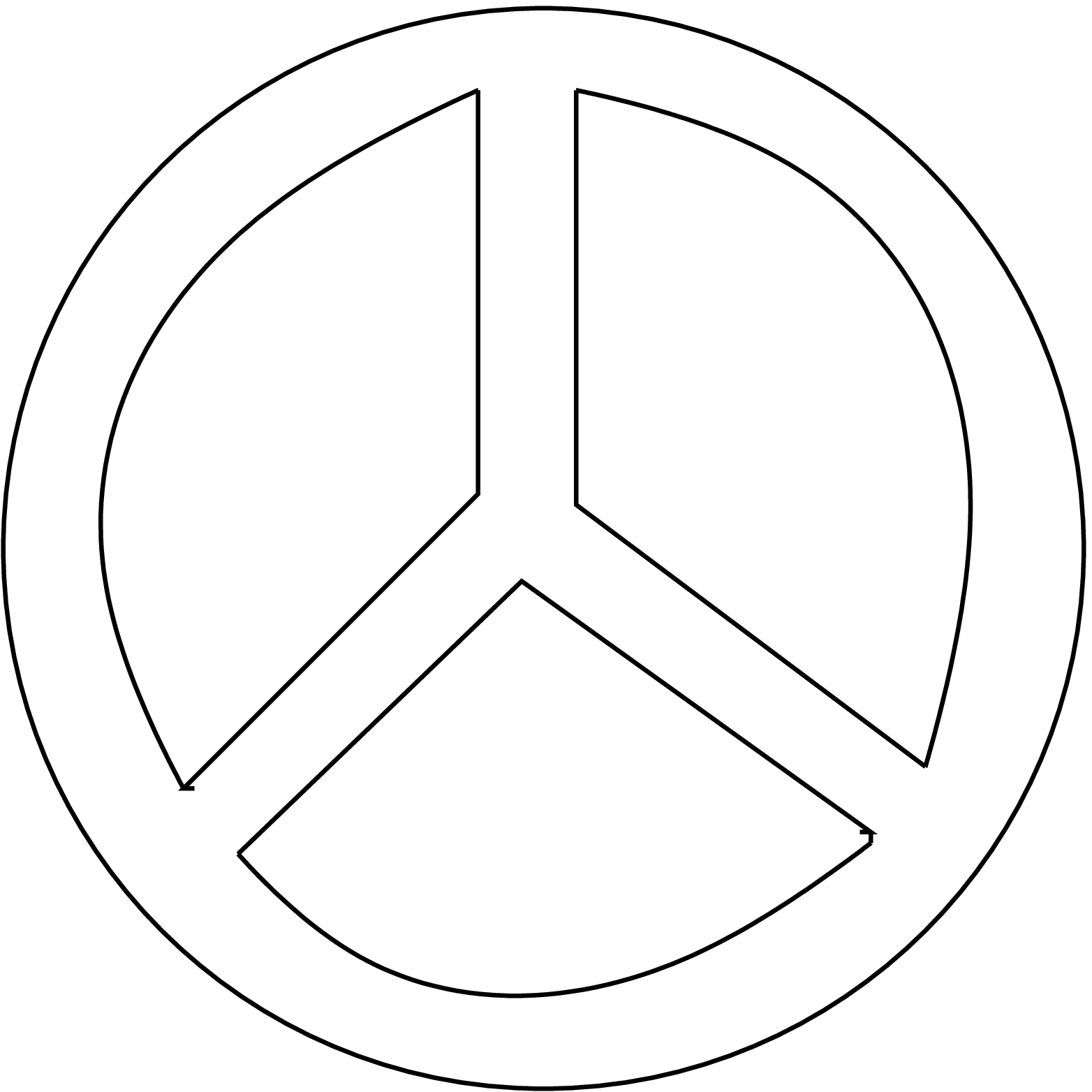,height=6cm,width=6cm}}
\caption{Ribbon graph for a sphere triangulated by four
triangles.} \label{one}
\end{figure}

\begin{figure}[h]
\centerline{\psfig{figure=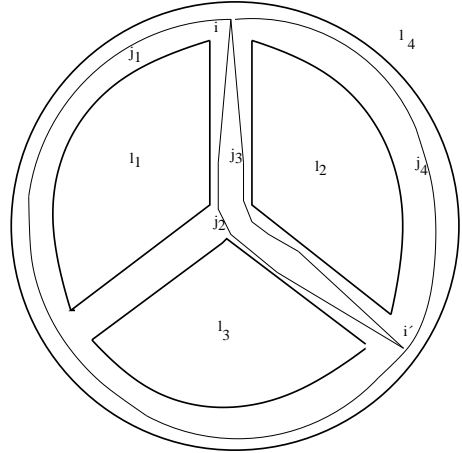,height=6cm,width=6cm}}
\caption{Colored ribbon graph for a sphere with an embedded
$\th_4$ spin network.} \label{two}
\end{figure}

\begin{figure}[h]
\centerline{\psfig{figure=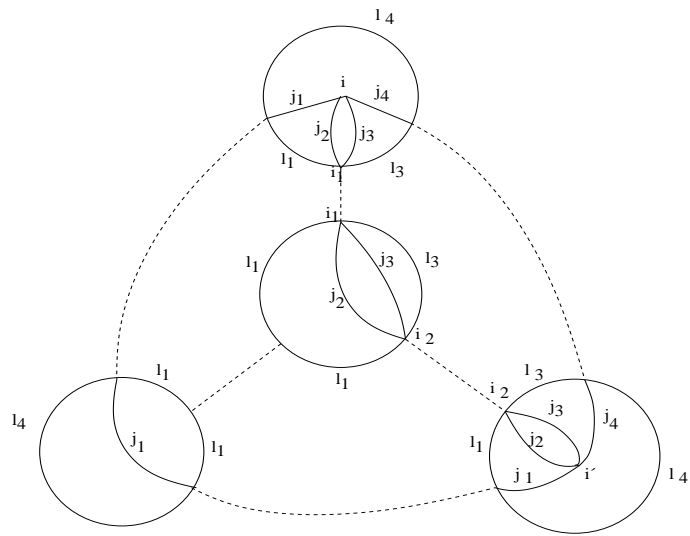,height=7cm,width=9cm}}
\caption{Vertex spin networks for the ribbon graph from Fig. 2.}
\label{three}
\end{figure}

\begin{figure}[h]
\centerline{\psfig{figure=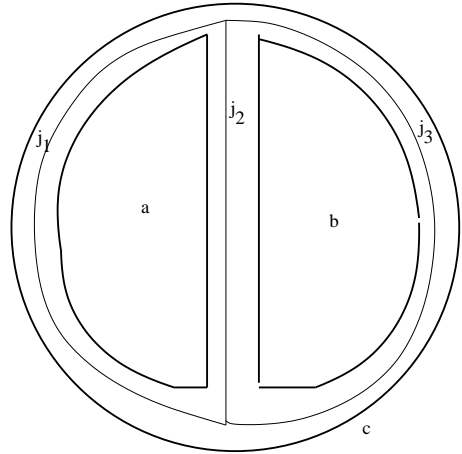,height=6cm,width=6cm}}
\caption{Ribbon graph for a sphere triangulated by two triangles
with an embedded $\th_3$ spin network.} \label{four}
\end{figure}

\begin{figure}[h]
\centerline{\psfig{figure=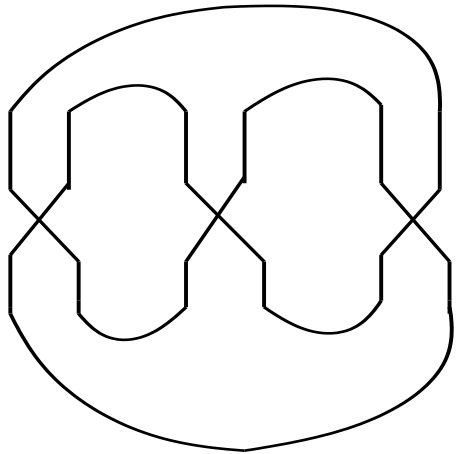,height=6cm,width=6cm}}
\caption{Ribbon graph for a torus triangulated by two triangles.}
\label{five}
\end{figure}

\begin{figure}[h]
\centerline{\psfig{figure=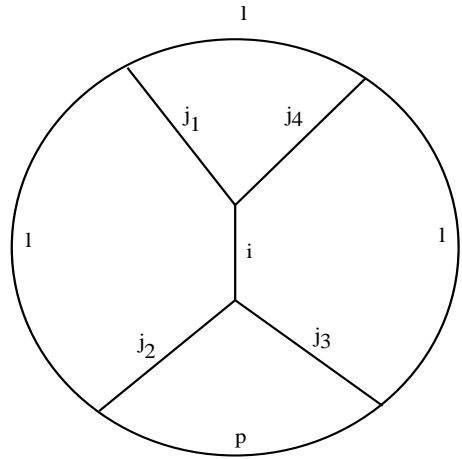,height=6cm,width=6cm}}
\caption{Vertex spin network for a $\th_4$ spin network embedded
in a torus.} \label{six}
\end{figure}

\begin{figure}[h]
\centerline{\psfig{figure=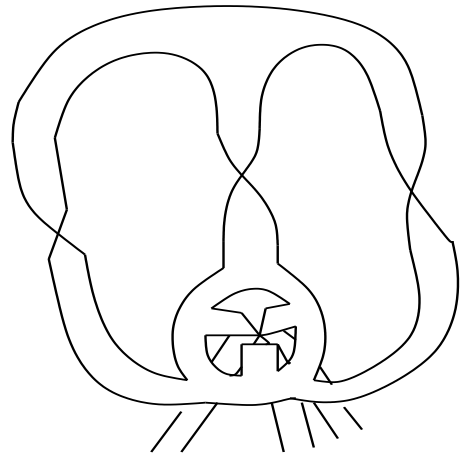,height=6cm,width=6cm}}
\caption{Ribbon graph for a triangulated torus with a hole.}
\label{seven}
\end{figure}

\begin{figure}[h]
\centerline{\psfig{figure=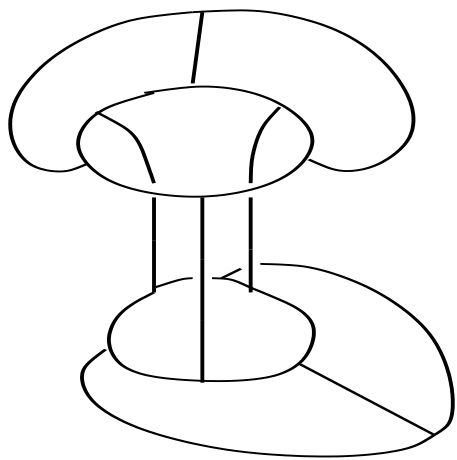,height=6cm,width=6cm}}
\caption{Dual one-complex for a triangulation of a genus two
surface.} \label{eight}
\end{figure}

\begin{figure}[h]
\centerline{\psfig{figure=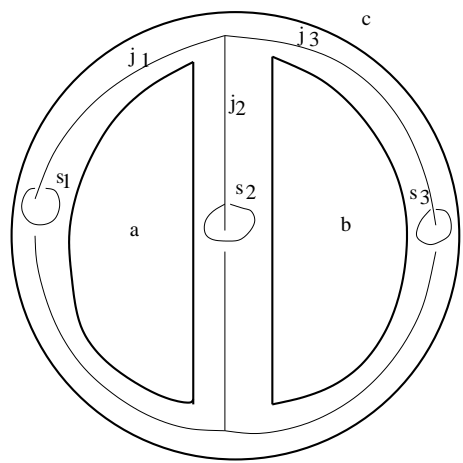,height=6cm,width=6cm}}
\caption{Ribbon graph for a sphere with an embedded theta spin
network linked with three Wilson loops.} \label{nine}
\end{figure}

\begin{figure}[h]
\centerline{\psfig{figure=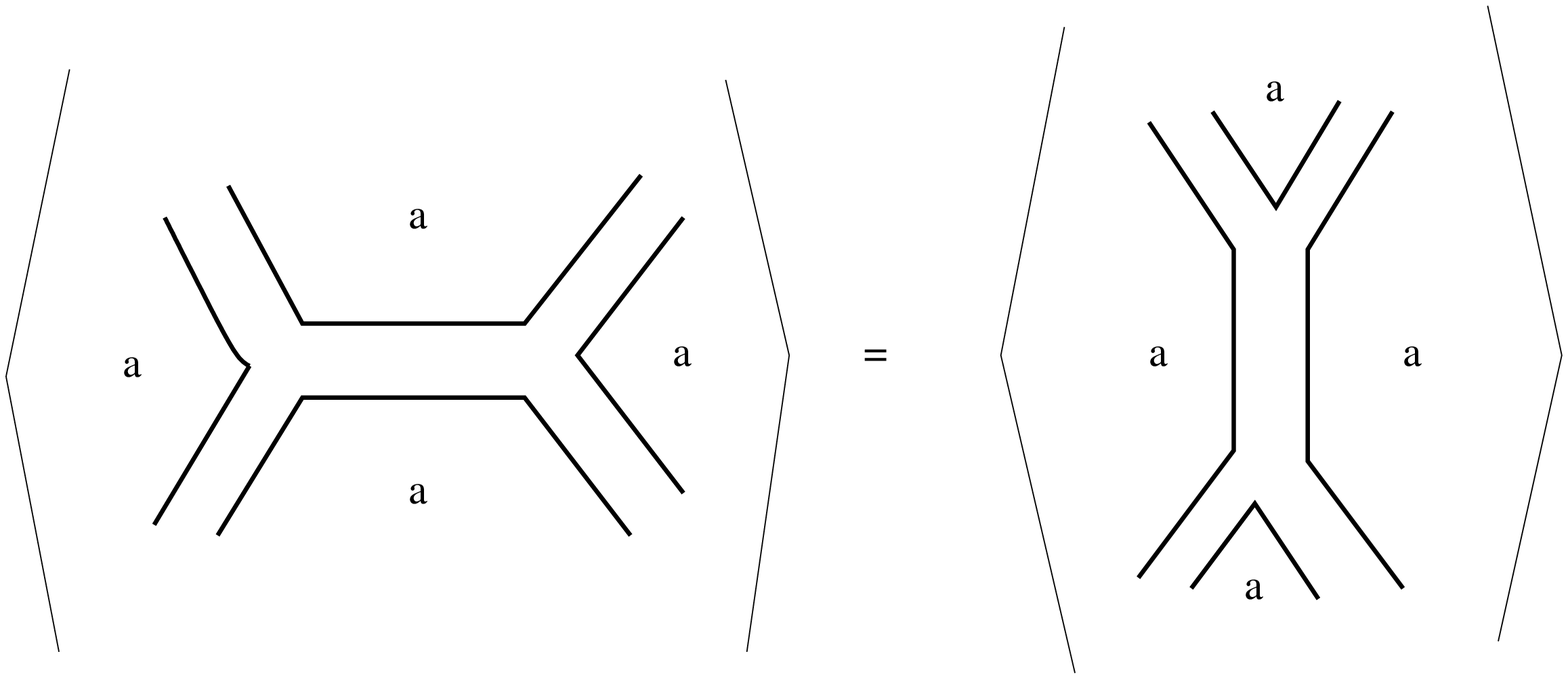,height=5cm,width=9cm}}
\caption{Invariance of the state sum under the $(2,2)$ Pachner
move which acts on a region where none of the edges of the $\th_n$
spin network pass.} \label{ten}
\end{figure}

\begin{figure}[h]
\centerline{\psfig{figure=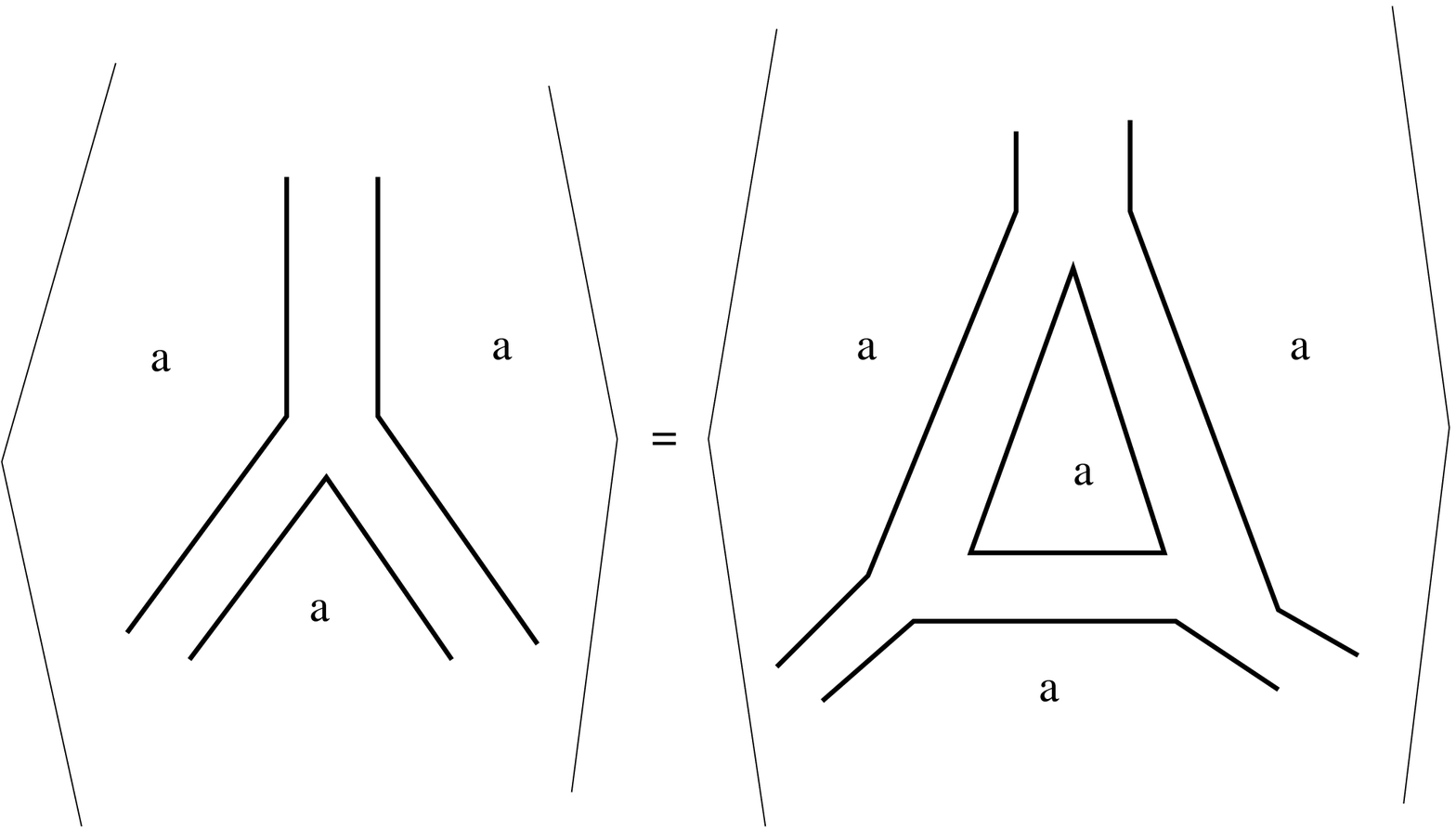,height=5cm,width=9cm}}
\caption{Invariance of the state sum under the $(1,3)$ Pachner
move which acts on a region where none of the edges of the $\th_n$
spin network pass.} \label{eleven}
\end{figure}

\begin{figure}[h]
\centerline{\psfig{figure=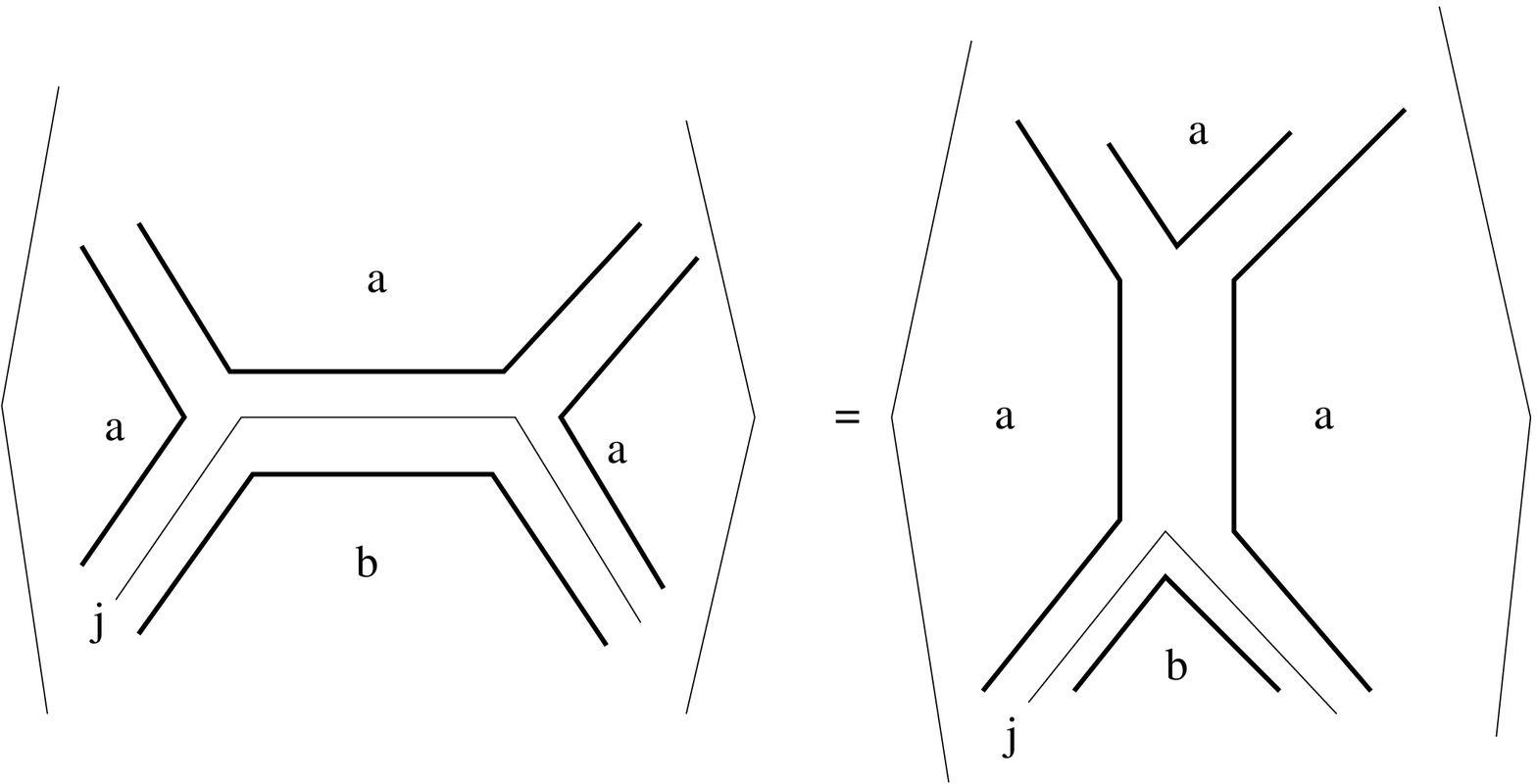,height=5cm,width=9cm}}
\caption{Invariance of the state sum under the $(2,2)$ Pachner
move which acts on a region where one of the $\th_n$ spin network
edges pass.} \label{twelve}
\end{figure}

\begin{figure}[h]
\centerline{\psfig{figure=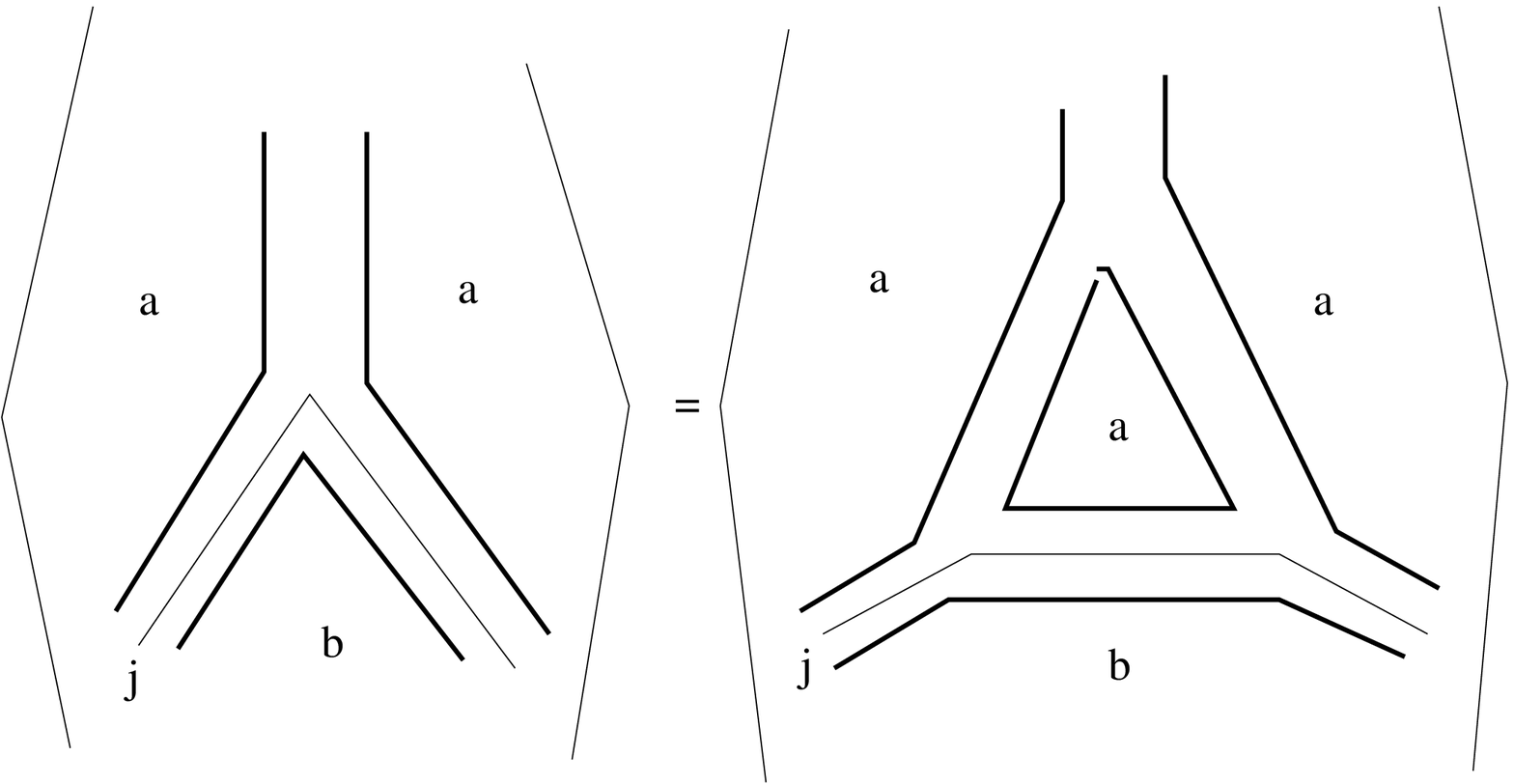,height=5cm,width=9cm}}
\caption{Invariance of the state sum under the $(1,3)$ Pachner
move which acts on a region where one of the $\th_n$ spin network
edges pass.} \label{thirteen}
\end{figure}

\begin{figure}[h]
\centerline{\psfig{figure=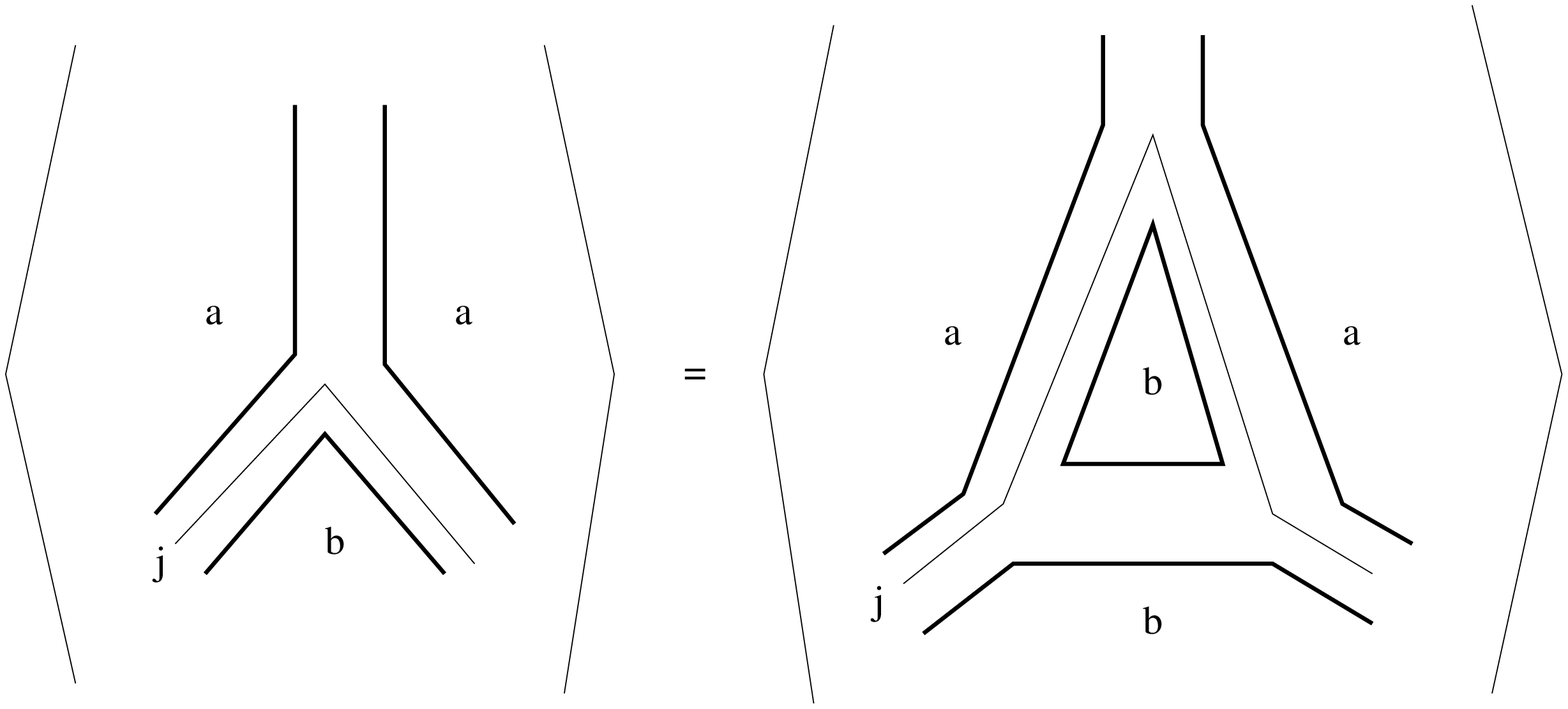,height=5cm,width=9cm}}
\caption{Another form of the invariance under the $(1,3)$ Pachner
move from Fig. 13.} \label{fourteen}
\end{figure}

\begin{figure}[h]
\centerline{\psfig{figure=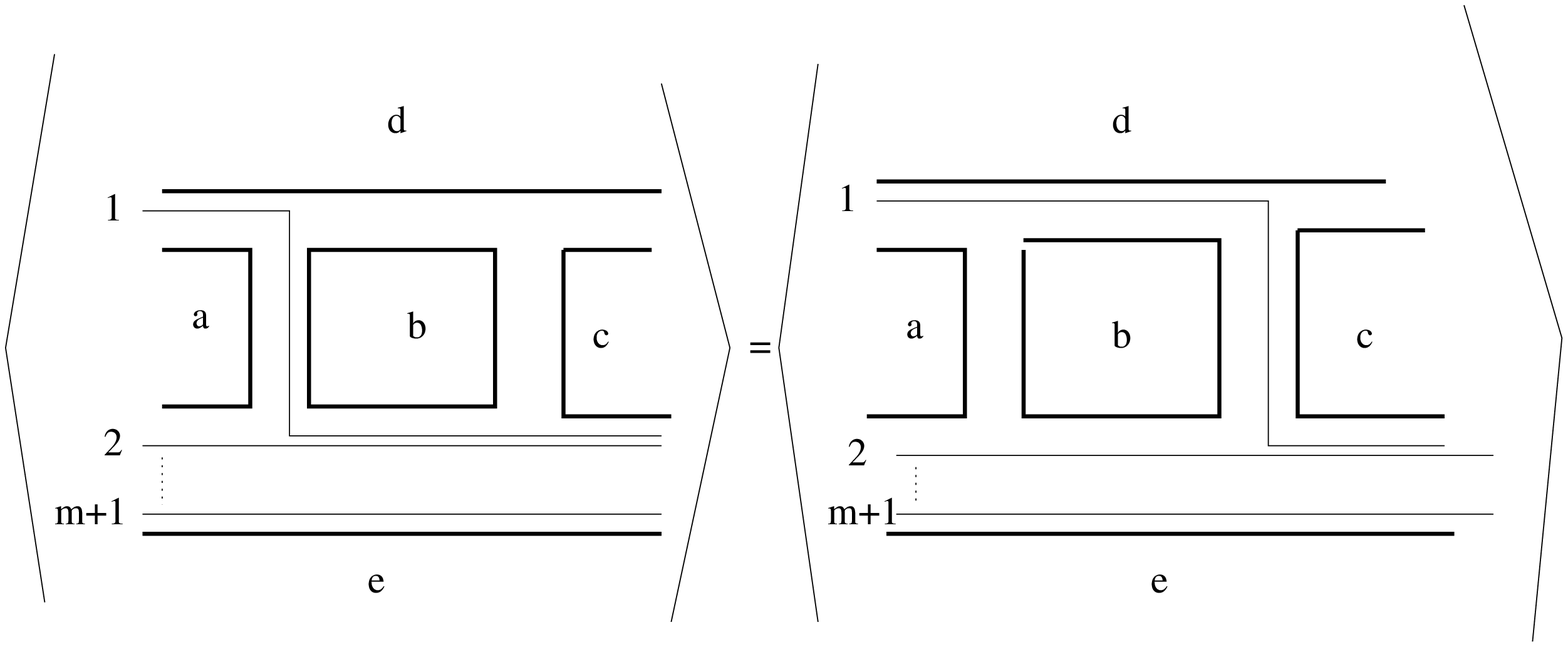,height=5cm,width=9cm}}
\caption{Invariance of the state sum under the elementary move of
one of the $\th_n$ spin network edges.} \label{fifteen}
\end{figure}

\begin{figure}[h]
\centerline{\psfig{figure=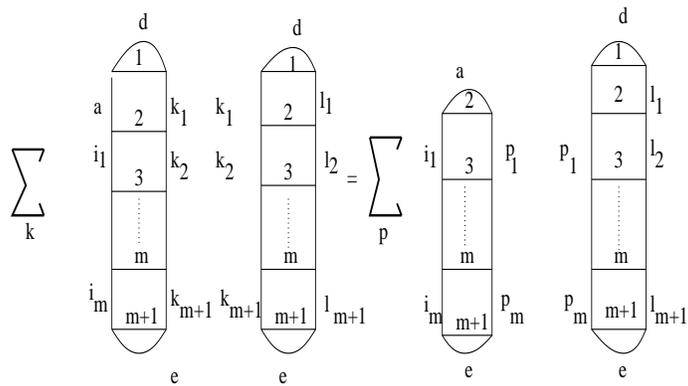,height=5cm,width=9cm}}
\caption{Spin network identity following from the elementary move
invariance from Fig. 15.} \label{sixteen}
\end{figure}

\end{document}